\documentclass{IEEEtran4P}
\ifCLASSINFOpdf
   \usepackage[pdftex]{graphicx}
\else
   \usepackage[dvips]{graphicx}
\fi
\usepackage[cmex10]{amsmath}
\usepackage[utf8]{inputenc}
\usepackage[english]{babel}
\usepackage{multirow}
\usepackage{graphicx}
\graphicspath{ {./images/} }
\usepackage{booktabs}
\usepackage{graphicx}
\usepackage{tabularx}
\usepackage{etoolbox}
\usepackage{nomencl}
\usepackage[linesnumbered,ruled,vlined]{algorithm2e}

\usepackage{lipsum} 

\usepackage{tikz}
\usepackage{array}
\makenomenclature
\usepackage{amsfonts}
\usepackage{adjustbox}
\usepackage{tabularray}
\usepackage{breqn}
\usepackage{color}
\usepackage{makecell}
\usepackage{amsmath}
\usepackage{nicematrix}
\usepackage{cite}
\usepackage{textcomp}
\usepackage{stfloats}
\usepackage{url}
\usepackage{verbatim}
\usepackage{cite}
\usepackage{relsize}
\usepackage[caption=false,font=normalsize,labelfont=sf,textfont=sf]{subfig}
\SetKwInput{KwInput}{Input}                
\SetKwInput{KwOutput}{Output}   
\usepackage{hyphenat}
\usepackage{comment}
\usepackage{multirow}
\usepackage{siunitx} 
\usepackage{booktabs} 
\usepackage{algpseudocode}
\usepackage{rotating}
\usepackage{threeparttable}
\usepackage{comment}
\newif\ifarxiv
\arxivfalse

\SetKwComment{Comment}{/* }{ */}

\usetikzlibrary{decorations.pathreplacing,calc}

\renewcommand{\baselinestretch}{1}
\begin{document}

\title{Optimizing Parameters of the DC Power Flow}

\author{Babak Taheri and Daniel K. Molzahn \\
School of Electrical and Computer Engineering\\
Georgia Institute of Technology\\
Atlanta, Georgia, USA\\
\texttt{\{taheri, molzahn\}@gatech.edu}
        }

\maketitle

\begin{abstract}
Many power system operation and planning problems use the DC power flow approximation to address computational challenges from the nonlinearity of the AC power flow equations. 
The DC power flow simplifies the AC power flow equations to a linear form that relates active power flows to phase angle differences across branches, parameterized by coefficients based on the branches' susceptances. 
Inspired by techniques for training machine learning models, this paper proposes an algorithm that seeks optimal coefficient and bias parameters to improve the DC power flow approximation's accuracy. 
Specifically, the proposed algorithm selects the coefficient and bias parameter values that minimize the discrepancy, across a specified set of operational scenarios, between the power flows given by the DC approximation and the power flows from the AC equations. Gradient-based optimization methods like Broyden-Fletcher-Goldfarb-Shanno (BFGS), Limited-Memory BFGS \mbox{(L-BFGS)}, and Truncated Newton Conjugate-Gradient (TNC) enable solution of the proposed algorithm for large systems.
After an off-line training phase, the optimized parameters are used to improve the accuracy of the DC power flow during on-line computations. Numerical results show several orders of magnitude improvements in accuracy relative to a hot-start DC power flow approximation across a range of test cases.%
\end{abstract}

\begin{IEEEkeywords}
DC power flow, machine learning, parameter optimization
\end{IEEEkeywords}

\thanksto{\noindent 
This research was supported by NSF award \#2145564.}

\vspace{-2em}
\section{Introduction}
\label{sec:Introduction}

Power flow analyses are integral to many applications, such as transfer capacity calculations, transmission loading relief, optimal dispatch, unit commitment, expansion planning, etc.~\cite{stott1974review, acc2016_tutorial}. By relating the voltage phasors, power injections, and power flows, the AC power flow equations accurately model the power grid for such applications. However, the nonlinearity of these equations introduces computational challenges such as non-convexities in optimization problems~\cite{bienstock2019strong, hiskens2001,overbye2004comparison}. These challenges frequently limit the direct use of the AC power flow equations in many applications, particularly those demanding large-scale and time-critical computations such as contingency analysis, day-ahead security-constrained unit commitment (SCUC), and real-time security-constrained  economic dispatch (SCED).

To bypass these challenges, engineers often resort to linear approximations~\cite{molzahn2019}. With a history dating back over a century~\cite{wilson1916}, the DC power flow approximation is one such widely used linearization, primarily due to its computational efficiency and its meaningful system representation. In the DC power flow approximation, the active power flow between buses $i$ and $j$, $p_{ij}$, is dictated by the phase angle difference $\theta_i - \theta_j$ with proportionality coefficient $b_{ij}$: $p_{ij} = b_{ij} (\theta_i - \theta_j)$.
The DC power flow plays an integral role across a broad range of applications, spanning both market operations and traditional power system operation and planning tasks~\cite{Stott2009}.

However, the selection of coefficients $b_{ij}$ in the DC power flow model, typically based on line parameters, often leaves room for improvement. With line resistance $r_{ij}$ and reactance $x_{ij}$, one might choose, for instance, $b_{ij} = 1/x_{ij}$ or $b_{ij} = -\Im(1/(r_{ij} + jx_{ij}))$, where $\Im(\,\cdot\,)$ takes the imaginary part of a complex argument. When $r_{ij} \neq 0$, these choices give similar but not the same values for $b_{ij}$. The choice of the $b_{ij}$ coefficients significantly impacts the DC power flow approximation's accuracy~\cite{Stott2009}. It is not always clear which coefficient choice provides the best accuracy for a specific application. There are also multiple ways to select bias parameters that adjust power injections and line flows to model shunts, HVDC infeeds, phase shifts, and line losses~\cite{Stott2009}.

This paper proposes a new approach for adaptively choosing the coefficients $b_{ij}$ and bias parameters in the DC power flow approximation. We leverage ideas from machine learning to tune these coefficients and biases, aiming to optimize the DC power flow approximation's accuracy. To accomplish this, an offline stage solves a plethora of AC power flow problems across a variety of operating conditions to construct a training dataset. Inspired by training methods for machine learning models, we then utilize a gradient-based method (e.g., Broyden-Fletcher-Goldfarb-Shanno (BFGS), Limited-Memory BFGS (L-BFGS), and Truncated Newton Conjugate-Gradient (TNC)) to optimize the values of the $b_{ij}$ and bias parameters. This process minimizes a specified loss function that quantifies the discrepancy between the DC power flow output and the AC power flow solutions over the training dataset. With these optimized parameter values, we can then apply the DC power flow approximation to the aforementioned applications to achieve accuracy improvements during online computations.

Our approach maintains the structure of the DC power flow approximation to enable seamless integration into many existing optimization models and computational algorithms that rely on this structure. We note that this is distinct from prior data-driven power flow modeling approaches, such as those proposed in~\cite{muhlpfordt2019optimal, zhentong2021, buason2022sample, chen2022, Liu2023, Shao2021}; see \cite{Jia203,jia2023tutorial1, jia2023tutorial2} for literature reviews. These prior approaches primarily focus on directly mapping power injections to power flows in a manner reminiscent of Power Transfer Distribution Factor (PTDF) models. These mappings typically disregard the underlying physical system topology, precluding physical intuition as operating conditions change. Moreover, they often consider quantities such as voltage magnitudes and reactive power injections that are neglected in typical DC power flow formulations. While these other power flow modeling approaches are useful in many settings, there is also substantial value in formulations that maintain the conventional DC power flow structure as this enables straightforward adoption in the many existing applications of DC power flow. For instance, using PTDF-style DC power flow models in optimal transmission switching problems (see, e.g.,~\cite{ruiz2017}) is substantially more complicated than DC power flow formulations that maintain phase angles~\cite{fisher2008}. Moreover, while a DC power flow approximation written in terms of phase angles can be easily translated to a PTDF formulation, the reverse is not straightforward, especially when the PTDF formulation is not based on the physical network structure, as is often the case in typical data-driven power flow models.



To summarize, the main contributions of this paper are:
\begin{enumerate}

    \item Introducing an optimization algorithm that adaptively selects the DC power flow approximation's coefficients $b_{ij}$ and bias parameters that adjust the power injections to account for bus shunt admittances, HVDC infeeds, phase shift injections, and branch losses.

    \item Utilizing and comparing various numerical methods such as BFGS, L-BFGS, and TNC to scale our proposed algorithm to large power systems.
    
    \item Providing numerical results that demonstrate the superior accuracy of our proposed algorithm under both normal and contingency conditions.
\end{enumerate}

The remainder of this paper is organized as follows: Section~\ref{sec:Power Flow Formulation} overviews the power flow equations. Section~\ref{sec:Optimizing Parameters} presents our proposed algorithm and solution method. Section~\ref{sec:Numerical Results} numerically demonstrates and benchmarks our proposed algorithm.  Section~\ref{sec:Conclusion} concludes the paper.


\section{Background on Power Flow Formulations}
\label{sec:Power Flow Formulation}

This section describes the AC power flow and the DC power flow approximation. The AC power flow accurately describes a system's steady-state behavior via nonlinear equations. The DC power flow linearly approximates these equations, thus improving tractability at the cost of introducing inaccuracies.

\subsection{AC Power Flow}


The AC power flow equations model a power system via nonlinear relationships among voltage magnitudes, phase angles, and complex power injections and flows. We first establish notation. Let $\mathcal{N}$, $ref$, and $\mathcal{E}$ denote the set of buses, reference bus, and set of lines, respectively. Each bus $i\in\mathcal{N}$ has a voltage phasor $V_i$ with phase angle $\theta_{i}$, a complex power injection $S_i$, and a shunt admittance $Y_i^S$.
Complex power flows into each terminal of each line $(i,j)\in\mathcal{E}$ are denoted as $S_{ij}$ and $S_{ji}$. Each line $(i,j)\in\mathcal{E}$ has a series admittance parameter $Y_{ij}$ and a shunt admittance parameter $Y_{ij}^{sh}$. The real and imaginary parts of a complex number are denoted as $\Re(\,\cdot\,)$ and $\Im(\,\cdot\,)$, respectively.  The transpose of a matrix is represented by $(\,\cdot\,)^T$. 
The power flow equations are:%
\begin{subequations}\label{eq:AC_PF}%
\begin{align}%
\raisetag{0.75em} P_i =& \sum_{(i,j) \in \mathcal{E}} p_{ij} + V_i^2 \Re(Y^{S}_i), ~~~
Q_i = \sum_{(i,j) \in \mathcal{E}} q_{ij} - V_i^2 \Im(Y^{S}_i), \\
p_{i j} =& V_i^2\left(\Re(Y_{ij})+\Re(Y_{ij}^{sh})\right)- V_i V_j \Re(Y_{ij}) \cos (\theta_{i}-\theta_{j})\nonumber
&\\ 
& \quad  - V_i V_j \Im(Y_{ij}) \sin (\theta_{i}-\theta_{j}), \\
q_{i j} =& -V_i^2\left(\Im(Y_{ij}) +\Im(Y_{ij}^{sh}) \right) - V_i V_j \Re(Y_{ij})  \sin (\theta_{i}-\theta_{j})\nonumber &\\
& \quad + V_i V_j\Im(Y_{ij})  \cos (\theta_{i}-\theta_{j}).
\end{align}
\end{subequations}


For each bus $i$, $P_i=\Re(S_i)$ and $Q_i = \Im(S_i)$ are the real and reactive power injections, respectively. For line $(i,j)\in\mathcal{E}$,  $p_{ij}$ and $q_{ij} $ are the real and reactive power flows, respectively. 

\subsection{DC Power Flow}
As discussed in~\cite{Stott2009}, there are two categories of DC power flow models, ``cold-start'' and ``hot-start'', that assume differing levels of information about the system's operating conditions. Here, we will introduce a generic formulation suitable for both cold-start and hot-start formulations and then show how these formulations differ in their parameter choices.

The DC power flow approximation uses several assumptions to linearize the non-linear AC power flow equations: neglect reactive power, assume all voltage magnitudes are constant, and consider angle differences across each transmission line to be small such that the small angle approximation for the sine function is applicable. Applying these assumptions to~\eqref{eq:AC_PF} yields the DC power flow:
%
%
\begin{subequations}
\label{eq:DCPF}
\begin{align}
P_i - \gamma_{i}&= \sum_{j \in \mathcal{N}} b_{ij} \cdot (\theta_i - \theta_j),
\label{eq:PowerBalance}\\
p^{DC}_{ij} &= b_{ij} \cdot (\theta_{i}-\theta_{j}) + \rho_{ij},
\label{eq:Line_flow}
\end{align}
\end{subequations}
where $p^{DC}_{ij}$ is the power flow in line $(i,j) \in \mathcal{E}$. As discussed in~\cite{Stott2009}, $\gamma_i$ is a bias parameter that accounts for losses from shunts, HVDC infeeds, and injections modeling phase shifts and branch losses for lines connected to bus~$i$. The bias parameter $\rho_{ij}$ for line $(i,j) \in \mathcal{E}$ is associated with line losses.


Let $\mathcal{N}^{\prime} = \mathcal{N} \setminus {ref}$ represent the set of all buses excluding the reference bus, $\mathbf{P}$ be the vector of net power injections at buses $i\in\mathcal{N}^{\prime}$, and $\boldsymbol{\theta}$ be the vector of voltage angles at buses $i\in\mathcal{N}^{\prime}$. Set $\theta_{ref} = 0$. Furthermore, define $\mathbf{A}$ as the $|\mathcal{E}| \times (|\mathcal{N}|-1)$ node-arc incidence matrix describing the connections between the system's buses and branches and let $\mathbf{b}$ be a length-$|\mathcal{E}|$ coefficient vector usually obtained using the branch susceptances.
The matrix form of~\eqref{eq:DCPF} is:
\begin{subequations}
\label{eq:DCPF_matrix}
\begin{align}
\mathbf{P} - \boldsymbol{\gamma}&= \mathbf{B}^{\prime} \cdot \boldsymbol{\theta},
\label{eq:MatrixForm}\\
\mathbf{p}^{DC} &= \left(\text{diag}(\mathbf{b}) \cdot \mathbf{A} \cdot \boldsymbol{\theta}\right) +\boldsymbol{\rho},
\label{eq:branchPowerFlow}
\intertext{where $\text{diag}(\,\cdot\,)$ is the diagonal matrix with the argument on the diagonal and $\mathbf{B}^{\prime}$ is}
\mathbf{B}^{\prime} &= \mathbf{A}^T \cdot \text{diag}(\mathbf{b}) \cdot \mathbf{A}. \label{eq:B-prime_sub}
\end{align}
\end{subequations}


%

\noindent In~\eqref{eq:DCPF_matrix}, $\mathbf{p}^{DC}$ is a length-$|\mathcal{E}|$ vector of power flows for each branch and $\boldsymbol{\rho}$ is a length-$|\mathcal{E}|$ vector associated with line losses. 

Solving \eqref{eq:MatrixForm} for $\boldsymbol{\theta}$ and substituting into \eqref{eq:branchPowerFlow} yields the so-called PTDF formulation of the DC power flow equations that linearly relates the line flows and real power injections:
\begin{equation}
\mathbf{p}^{DC} = \text{diag}(\mathbf{b}) \cdot \mathbf{A} \cdot [\mathbf{A}^T \cdot \text{diag}(\mathbf{b}) \cdot \mathbf{A}]^{-1}\cdot (\mathbf{P} - \boldsymbol{\gamma}) + \boldsymbol{\rho}.
\label{eq:branchPowerFlow2}
\end{equation}

The parameters $\mathbf{b}$, $\boldsymbol{\gamma}$, and $\boldsymbol{\rho}$ impact the DC power flow's performance. Cold- and hot-start versions of the DC power flow assume different amounts of prior information when choosing these parameters.

\subsubsection{Cold-start DC power flow}
In this version, the coefficient and bias parameters are selected without relying on a
nominal AC power flow solution. For instance, the coefficient $b_{ij}$ can be selected as either:
\begin{equation}
b^{cold}_{ij} = \Im\left(\frac{-1}{r_{ij} + j \cdot x_{ij}}\right) \text{~~~~or~~~~}
b^{cold}_{r=0,ij} = \frac{1}{x_{ij}}.
\end{equation}

Furthermore, the bias parameters ($\boldsymbol{\gamma}$ and $\boldsymbol{\rho}$) are typically set to zero in the cold-start version. These heuristic methods offer simplicity but may not provide adequate accuracy.


\subsubsection{Hot-start DC power flow}
 A nominal AC power flow solution can provide a good starting point to construct a DC power flow approximation \cite{Stott2009}. For instance, the so-called ``localized loss modeling'' variant of the hot-start DC model in~\cite{Stott2009} selects:
\begin{subequations}
    \begin{align}
        b^{hot}_{ij}&=b_{ij}  v^{\bullet}_{i}v^{\bullet}_{j} \sin(\theta^{\bullet}_{i}-\theta^{\bullet}_{j})/({\theta^{\bullet}_{i}-\theta^{\bullet}_{j}}), \label{eq:sub1-hot-start} \\
        \gamma^{hot}_{i} &= \sum_{(i,j) \in \mathcal{E}} \Re(Y_{ij}) v^{\bullet}_{i}(v^{\bullet}_{i}-v^{\bullet}_{j} \cos(\theta^{\bullet}_{i}-\theta^{\bullet}_{j})), \label{eq:sub2-hot-start}\\
        \rho^{hot}_{ij}&= \Re(Y_{ij}) v^{\bullet}_{i}(v^{\bullet}_{i}-v^{\bullet}_{j} \cos(\theta^{\bullet}_{i}-\theta^{\bullet}_{j})), \label{eq:sub3-hot-start}
    \end{align}%
    \label{eq:hot-start}%
\end{subequations}%
where $(\,\cdot\,)^{\bullet}$ denotes quantities from the nominal AC power flow solution. Note that $\boldsymbol{\gamma}^{hot}$ denotes injections that model the impacts of branch losses on phase angles and $\boldsymbol{\rho}^{hot}$ accounts for the branch losses in the line flow expressions themselves.

In the next section, we introduce a machine learning-inspired algorithm to optimize the coefficient ($\mathbf{b}$) and bias ($\boldsymbol{\gamma}$ and $\boldsymbol{\rho}$) parameters in the DC power flow model. Our proposed algorithm aims to reduce the discrepancy between the power flows predicted by the DC power flow model and the actual power flows from the AC power flow equations.


\section{Parameter Optimization Algorithm}
\label{sec:Optimizing Parameters}

\begin{figure}[t]
    \centering
    \includegraphics[width=3.5in,trim={0 1.75em 0 0.5em},clip]{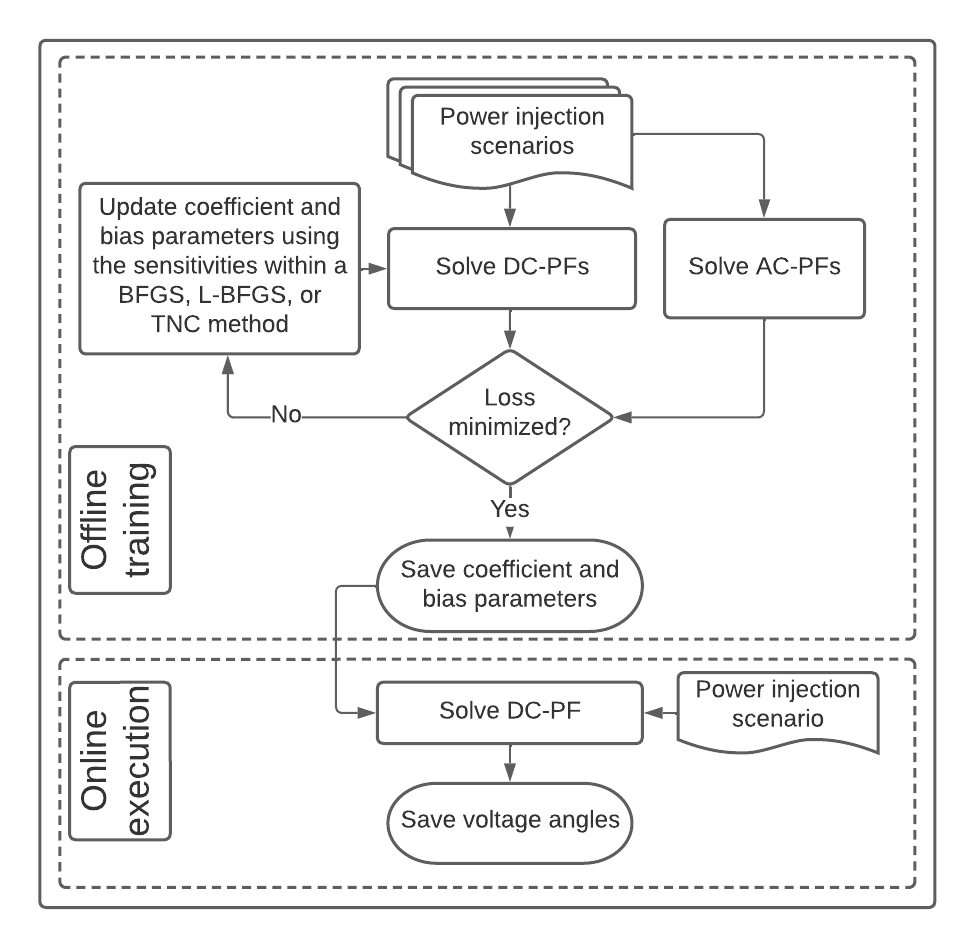}
    \caption{Flowchart describing the proposed algorithm.}
    \label{fig:flowchart}
    \vspace{-1.5em}
\end{figure}

As illustrated in Fig.~\ref{fig:flowchart}, our parameter optimization algorithm consists of \textit{offline} and \textit{online} stages. The \textit{offline} stage, a one-time procedure, focuses on computing the optimal parameters, $\mathbf{b}$, $\boldsymbol{\gamma}$, and $\boldsymbol{\rho}$, over a range of power injection scenarios. This ensures that our DC model closely aligns with the AC power flow across diverse operating conditions. In the \textit{online} phase, the DC model, equipped with these optimized parameters, offers rapid and accurate approximations suitable for real-time tasks. Thus, our algorithm invests computational time upfront during offline optimization to reap continual benefits during online applications.
Our algorithm is inspired by supervised machine learning: we use power injections as inputs and line flows as targets. However, no traditional machine learning models or neural networks are applied.
%
Rather, the offline phase refines the parameters $\mathbf{b}$, $\boldsymbol{\gamma}$, and $\boldsymbol{\rho}$ by solving an optimization problem which has these parameters as decision variables.
%

To optimize the parameters $\mathbf{b}$, $\boldsymbol{\gamma}$, and $\boldsymbol{\rho}$, we first formulate a loss function that quantifies the accuracy of the DC power flow approximation by comparing the DC approximation's power flow predictions against the power flows obtained from the AC power flow equations across a set of power injection scenarios.

We next compute the sensitivities of the loss function with respect to these parameters.
These sensitivities guide the optimization process by indicating the direction in which the parameters should be adjusted to minimize the loss function. Using this sensitivity information, we then apply an optimization method, such as BFGS, L-BFGS, and TNC. These methods offer scalable optimization capabilities, making them well-suited for large power systems. By optimally selecting the parameter values using our algorithm, the DC power flow model's accuracy can be significantly improved across a broad range of power injection scenarios. The details of this algorithm and its implications are presented next.

\subsection{Loss Function}
\label{subsec:Loss Function}
Here, we introduce a loss function based on the sum of squared two-norm discrepancies between the AC ($\mathbf{p}_{m}^{AC}$) and DC ($\mathbf{p}_{m}^{DC}$) power flow models across a specified set of power injection scenarios $\mathcal{M} = {1, 2, \ldots, S}$. This approach is typical in machine learning for its robustness and differentiability. Our loss function, $\mathcal{L}$, is formulated as:
\begin{align}
\nonumber \mathcal{L}(\mathbf{b}, \boldsymbol{\gamma}, \boldsymbol{\rho}) &= \frac{1}{|\mathcal{E}|}\sum_{m \in \mathcal{M}} ||\mathbf{p}_{m}^{DC} - \mathbf{p}_{m}^{AC}||^2_2,\\ 
& = \frac{1}{|\mathcal{E}|}\sum_{m \in \mathcal{M}} ||\text{diag}(\mathbf{b}) \cdot \mathbf{A} \cdot [\mathbf{A}^T \cdot \text{diag}(\mathbf{b})\mathbf{A}]^{-1} \nonumber\\ \label{eq:objective_function}%
& \qquad \qquad \qquad \times  (\mathbf{P}_{m} - \boldsymbol{\gamma}) +\boldsymbol{\rho} - \mathbf{p}_{m}^{AC}||^2_2, 
\end{align}%
%
where the constant $\frac{1}{|\mathcal{E}|}$ normalizes this function based on the system size. As shown in \eqref{eq:branchPowerFlow2} and \eqref{eq:objective_function}, $\mathbf{p}_m^{DC}$ (and thus $\mathcal{L}(\mathbf{b}, \boldsymbol{\gamma},\boldsymbol{\rho})$) is a function of the coefficient parameters $\mathbf{b}$ and the bias parameters $\boldsymbol{\gamma}$ and $\boldsymbol{\rho}$.
By focusing on the two-norm of the discrepancies, larger deviations are penalized more heavily. This is well aligned with typical applications where a small number of severe approximation errors would be more problematic than a large number of minor errors. One could instead use other norms such as $L_{1}$ or $L_{\infty}$ without major conceptual changes.

Finally, the unconstrained optimization problem to find the best coefficient and bias parameters is formulated as:
\begin{equation}
    \min_{\mathbf{b},\boldsymbol{\gamma},\boldsymbol{\rho}} \quad \mathcal{L}(\mathbf{b},\boldsymbol{\gamma},\boldsymbol{\rho}).
    \label{eq:optimization problem}
\end{equation}

\subsection{Sensitivities of the Coefficient and Bias Parameters}
\label{subsec:Sensitivity Analysis}

Optimization methods such as BFGS, \mbox{L-BFGS}, and TNC rely on the gradient of the loss function with respect to the parameters in the $\mathbf{b}$, $\boldsymbol{\gamma}$, and $\boldsymbol{\rho}$ vectors, i.e., the sensitivity of $\mathcal{L}(\mathbf{b},\boldsymbol{\gamma}. \boldsymbol{\rho})$ to infinitesimal changes in $\mathbf{b}$, $\boldsymbol{\gamma}$, and $\boldsymbol{\rho}$ across all power injection scenarios. We first focus on sensitivities for the $\mathbf{b}$ parameters, denoted as $\mathbf{g}^{b}$, which are calculated by taking the derivatives of the loss function \eqref{eq:objective_function} with respect to the coefficient parameters $\mathbf{b}$:
\begin{subequations}
\label{eq:sensitivity}
\begin{equation}
\label{eq:sensitivity1}
 \mathbf{g}^{b}= \frac{2}{|\mathcal{E}|}\sum_{m \in \mathcal{M}} \left. \frac{\partial \mathbf{p}^{DC}}{\partial \mathbf{b}} \right|_ {\mathbf{p}^{DC}_{m}}\Big(\mathbf{p}_{m}^{DC} - \mathbf{p}_{m}^{AC} \Big),
\end{equation}
where $\frac{\partial \mathbf{p}^{DC}}{\partial \mathbf{b}}$ is obtained from the derivative of \eqref{eq:branchPowerFlow2} with respect to the coefficient parameters $\mathbf{b}$:
\begin{align}
\frac{\partial \mathbf{p}^{DC}}{\partial \mathbf{b}} = & \text{~diag} \Bigg(\Big [\mathbf{A} [\mathbf{A}^T \cdot \text{diag}(\mathbf{b}) \cdot \mathbf{A}]^{-1} (\mathbf{P}-\boldsymbol{\gamma}) \Big]^{T} \Bigg) \times \nonumber \\
&  \Big (\mathbf{I} - \text{diag}(\mathbf{b}) \cdot \mathbf{A} [\mathbf{A}^T \cdot \text{diag}(\mathbf{b}) \cdot \mathbf{A}]^{-1} \mathbf{A}^T \Big).
\label{eq:coef_sensitivity}
\end{align}
\end{subequations}
where $\mathbf{I}$ is the identity matrix. The appendix provides a detailed derivation of~\eqref{eq:coef_sensitivity}.

Like the coefficient parameters $\mathbf{b}$, the bias parameters $\boldsymbol{\gamma}$ significantly impact the accuracy of DC power flow. 
The gradient of the loss function with respect to the bias parameters $\boldsymbol{\gamma}$ is represented by $\mathbf{g}^{\gamma}$:
\begin{subequations}
\begin{equation}
\label{eq:sensitivity-bias}
 \mathbf{g}^{\gamma}= \frac{2}{|\mathcal{E}|}\sum_{m \in \mathcal{M}} \left. \frac{\partial \mathbf{p}^{DC}}{\partial \boldsymbol{\gamma}} \right|_ {\mathbf{p}^{DC}_{m}}\Big(\mathbf{p}_{m}^{DC} - \mathbf{p}_{m}^{AC} \Big),
\end{equation}
where $\frac{\partial \mathbf{p}^{DC}}{\partial \boldsymbol{\gamma}}$ is calculated by taking the derivative of \eqref{eq:branchPowerFlow2} with respect to bias parameters $\boldsymbol{\gamma}$:

\begin{align}
\frac{\partial \mathbf{p}^{DC}}{\partial \boldsymbol{\gamma}} = -\text{diag}(\mathbf{b}) \cdot \mathbf{A} [\mathbf{A}^T \cdot \text{diag}(\mathbf{b}) \cdot \mathbf{A}]^{-1}.
\label{eq:coef_sensitivity2}
\end{align}

\end{subequations}

Finally, the gradient of the loss function with respect to the bias parameters $\boldsymbol{\rho}$ is represented by $\mathbf{g}^{\rho}$:

\begin{equation}
\label{eq:sensitivity-rho}
 \mathbf{g}^{\rho}= \frac{2}{|\mathcal{E}|}\sum_{m \in \mathcal{M}} \left. \frac{\partial \mathbf{p}^{DC}}{\partial \boldsymbol{\rho}} \right|_ {\mathbf{p}^{DC}_{m}}\Big(\mathbf{p}_{m}^{DC} - \mathbf{p}_{m}^{AC} \Big),
\end{equation}
where $\frac{\partial \mathbf{p}^{DC}}{\partial \boldsymbol{\rho}}$ is calculated by taking the derivative of \eqref{eq:branchPowerFlow2} with respect to bias parameters $\boldsymbol{\rho}$, which is the identity matrix $\mathbf{I}$.

These sensitivities enable gradient-based methods for optimizing the parameters $\mathbf{b}$, $\boldsymbol{\gamma}$, and $\boldsymbol{\rho}$, as we will describe next.

\subsection{Optimization Formulation and Solution Methods}
\label{sec:Optimization Methods}
With known sensitivities, many gradient-based methods such as BFGS, L-BFGS, and TNC can be applied to the unconstrained optimization problem~\eqref{eq:optimization problem}. We next summarize the key characteristics of each method. Our numerical results in the following section empirically compare the performance of each method for a range of test cases. 

\textbf{BFGS}: An iterative quasi-Newton approach proposed by Broyden, Fletcher, Goldfarb, and Shanno~\cite[p.~136]{nocedal2006numerical}, BFGS uses the gradient to update an inverse Hessian matrix approximation, bypassing the need for the complete Hessian matrix.

\textbf{L-BFGS}: An evolution of the BFGS method that uses a limited memory approach to handle large datasets.

\textbf{Conjugate-Gradient (CG)}: The CG method  uses a nonlinear conjugate gradient algorithm~\cite[pp.~120-122]{nocedal2006numerical}, which only relies on the first derivatives.

\textbf{Newton-CG}: The Newton-CG method (also known as the truncated Newton method) uses a CG method to compute the search direction~\cite[p.~168]{nocedal2006numerical}.

\textbf{Truncated Newton Conjugate-Gradient (TNC)}: The TNC method uses a truncated Newton algorithm to minimize a function with variables subject to bounds~\cite{nocedal2006numerical, nash1984newton}.

We will numerically assess the performance of each optimization method when solving problem~\eqref{eq:optimization problem}.

\section{Experimental Results and Discussion}
\label{sec:Numerical Results}

This section presents and benchmarks the results obtained from our proposed algorithm. To demonstrate the model's efficacy, we compare power flows from our machine learning inspired algorithm to those from traditional DC power flow formulations and the AC power flow model. These comparisons consider multiple illustrative test systems from~\cite{birchfield2016grid, pglib}.

For these test cases, we generated $10,000$ power injection scenarios ($8,000$ for offline training and $2,000$ for testing). These scenarios were created by multiplying the nominal power injections by a normally distributed random
variable with mean of one and standard deviation of $10\%$. We initialize the proposed algorithm with hot-start parameters.
Solutions to the AC power flow problems were computed using \texttt{PowerModels.jl}~\cite{coffrin2018powermodels} on a computing node of the Partnership for an Advanced Computing Environment (PACE) cluster at Georgia Tech. This computing node has a 24-core CPU and 32~GB of RAM. The proposed algorithm is implemented in Python 3.0 using a Jupyter Notebook. To minimize the loss function, we used the BFGS, \mbox{L-BFGS}, TNC, CG, and Newton-CG implementations from the \texttt{scipy.optimize.minimize} library.


\subsection{Benchmarking Optimization Methods}
\begin{figure*}[t!]
\centering
\subfloat[\small Training loss ]{
\includegraphics[width=0.485\textwidth]{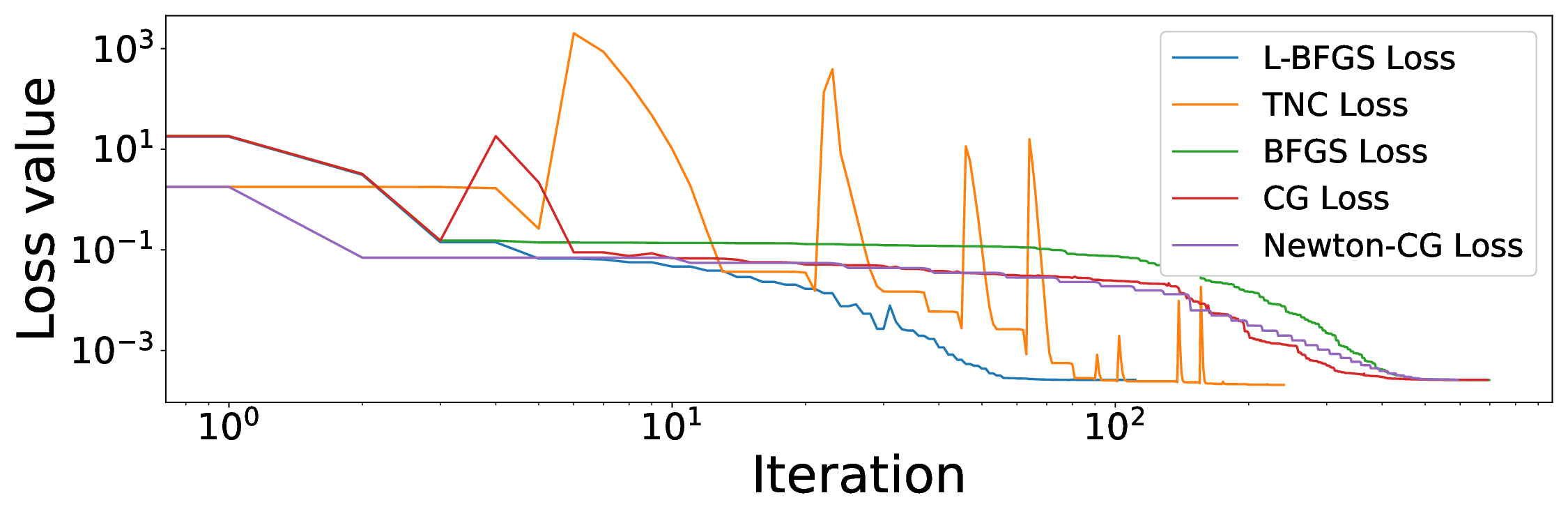}
\label{fig:300bus_Optimizer_Training_loss}
}
\hfill
\subfloat[\small  Training time]{
\includegraphics[width=0.485\textwidth]{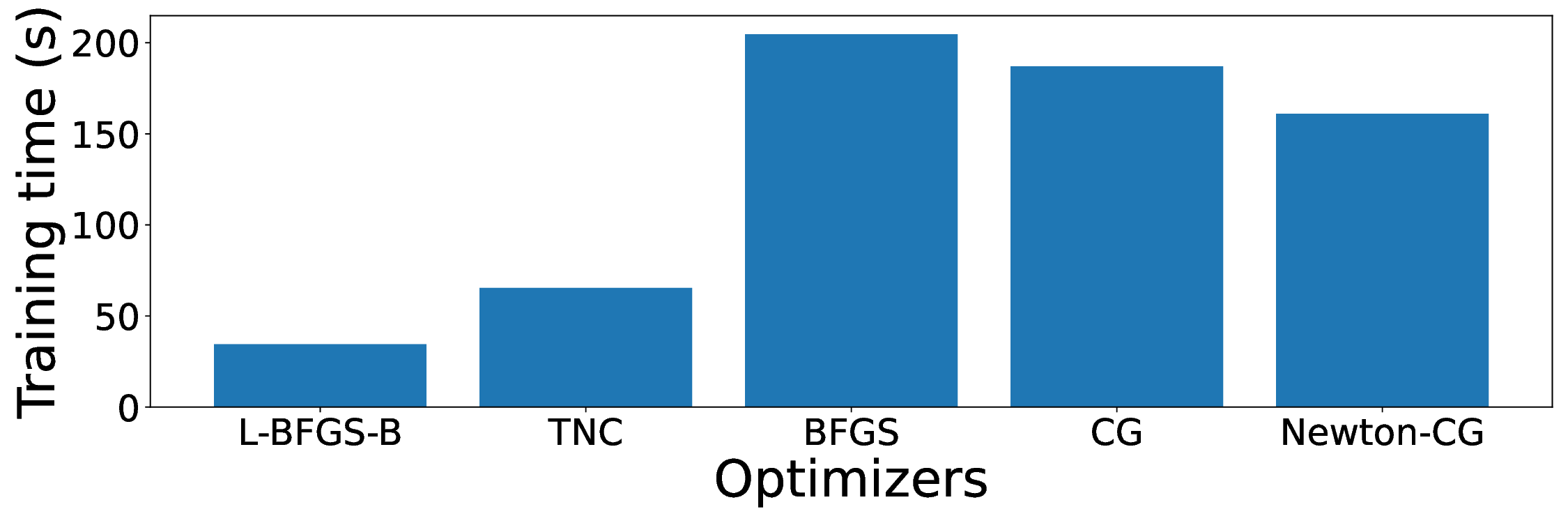}
\label{fig:300bus_Optimizer_Training_Times}
}
\caption{Training losses and times for the L-BFGS, TNC, BFGS, CG, and Newton-CG methods for the \texttt{IEEE 300-bus} system.}
\label{fig:training}
\end{figure*}

First, we assess the performance of the BFGS, L-BFGS, TNC, CG, and Newton-CG methods, as detailed in Section~\ref{sec:Optimization Methods}, using the \texttt{IEEE 300-bus} system as a representative example. Each method uses a convergence tolerance of $1\times10^{-6}$.
Fig.~\ref{fig:training} shows the evolution of the training loss (i.e., \eqref{eq:objective_function} evaluated for the training scenarios) and the training time for each method.
For the \texttt{IEEE 300-bus} system as well as the other test cases we considered, the L-BFGS method had the fastest performance during the offline training step for most of the cases. However, the quality of the resulting parameters, as measured by the loss function value for the training dataset, exhibited mixed results with the TNC, BFGS, \mbox{L-BFGS}, CG, and Newton-CG methods each achieving the best performance for some test cases. Due to their overall performance, we choose to focus on applying the TNC, BFGS, and L-BFGS methods to~\eqref{eq:optimization problem}. We will evaluate their performance on the test datasets in Section \ref{subsec:Comparative Analysis}. 

\subsection{Comparison of Parameter Values Across  Selection Methods}
Here, we illustrate the distributions of cold-start, hot-start, and optimized parameters $\mathbf{b}$, $\boldsymbol{\gamma}$, and $\boldsymbol{\rho}$ across various test systems using box plots. Each box shows the interquartile range, representing the middle $50\%$ of the data, with the central line indicating the median. The whiskers, extending from each box, display the data within $1.5$ times the interquartile range. Data points outside of these whiskers are considered outliers and are plotted as individual dots. The horizontal lines at the whiskers' ends indicate the $90^{th}$ percentile of the data. For each test system, the box plot figures highlight four distributions: the cold-start ($\mathbf{b}^{cold}$ or $\mathbf{b}^{cold}_{r=0}$), hot-start ($\mathbf{b}^{hot}$, $\boldsymbol{\gamma}^{hot}$, and $\boldsymbol{\rho}^{hot}$), and the results from our optimization algorithm ($\mathbf{b}^{opt}$, $\boldsymbol{\gamma}^{opt}$, and $\boldsymbol{\rho}^{opt}$). All data are visualized on a logarithmic scale, emphasizing variations across multiple orders of magnitude.

\begin{figure*}[t!]
\centering
\subfloat[\small ]{
\includegraphics[width=0.485\textwidth]{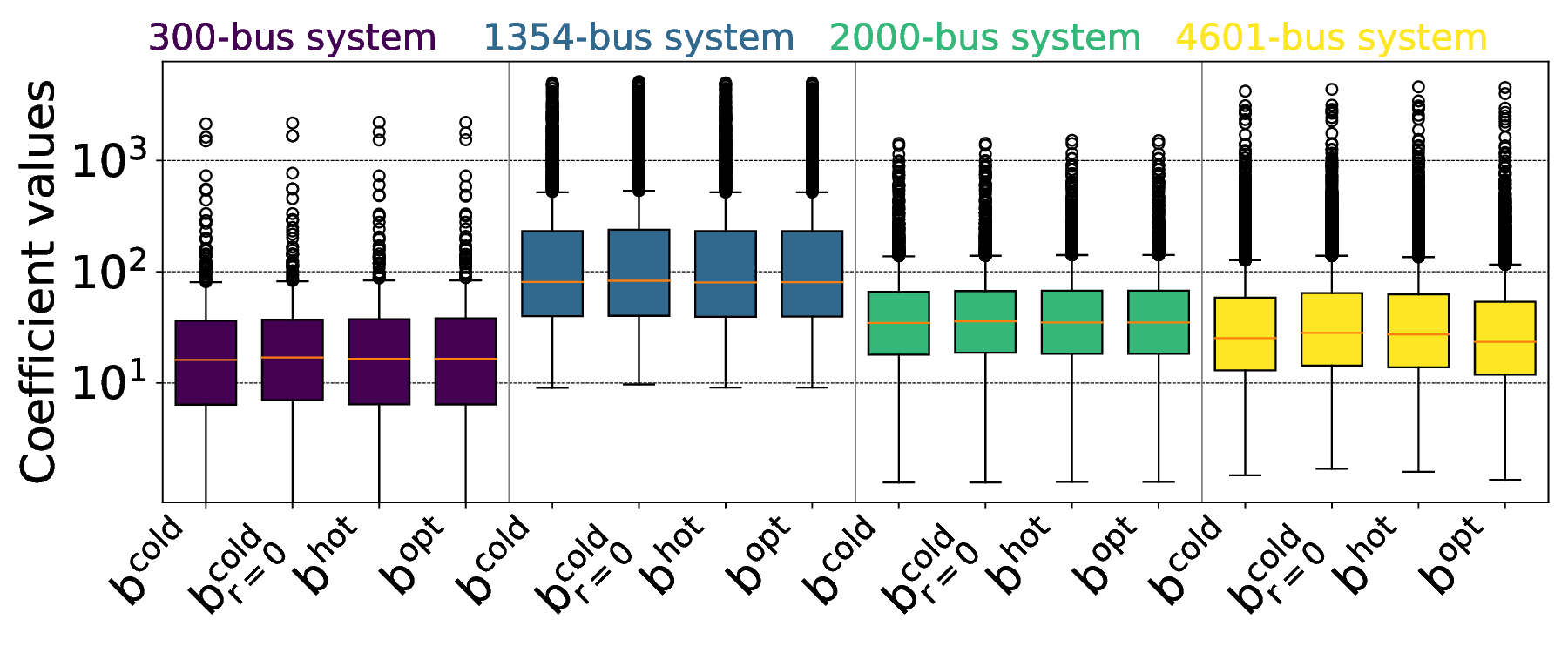}
    \label{fig:box-plot}
}
\hfill
\subfloat[\small  ]{
\includegraphics[width=0.485\textwidth]{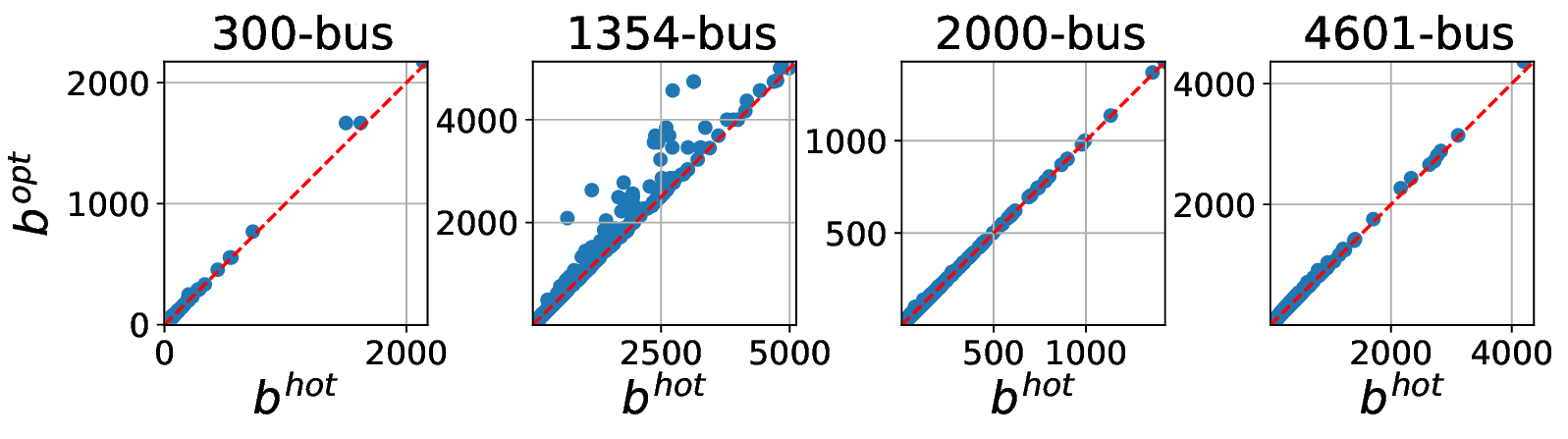}
    \label{fig:coefficient-scatter}
}
\caption{(a) Boxplots showing the distributions of the $\mathbf{b}$ parameter values for multiple test cases. Each test case is represented by four boxplots indicating the cold-start, hot-start, and the optimal $\mathbf{b}$ parameter values. (b) Scatter plots comparing the coefficient values  $\mathbf{b}^{hot}$ and $\mathbf{b}^{opt}$ for various test cases.}
\label{fig:Coeff}
\end{figure*}

\begin{figure*}[t!]
\centering
\subfloat[\small ]{
\includegraphics[width=0.485\textwidth]{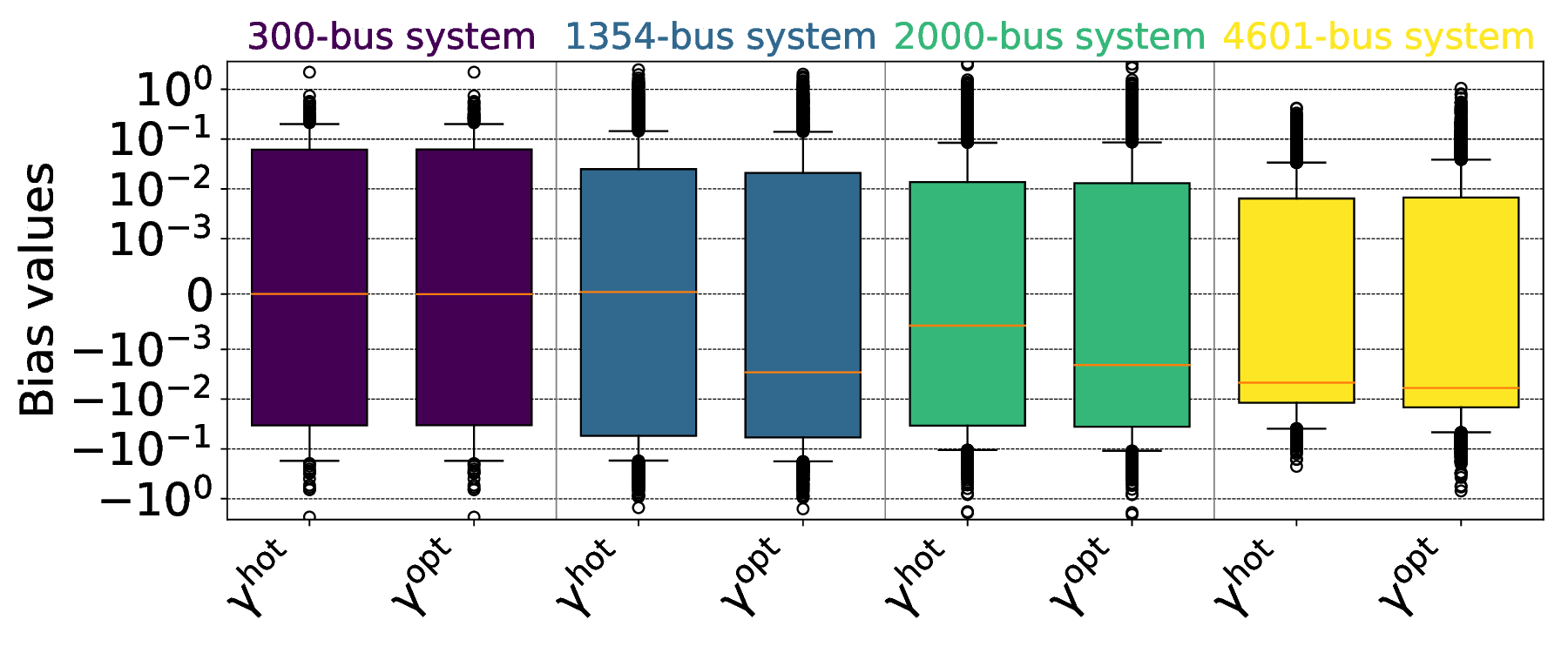}
    \label{fig:boxplot-biases}
}
\hfill
\subfloat[\small  ]{
\includegraphics[width=0.485\textwidth]
{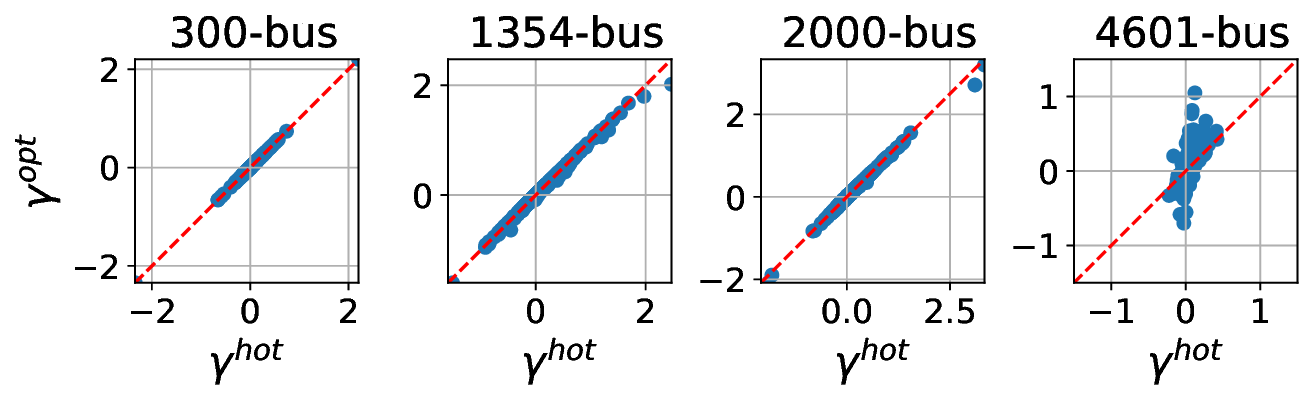}
    \label{fig:biases-scatter}
}
\caption{ a)~Boxplots showing the distributions of the hot-start and optimal injection bias values, $\boldsymbol{\gamma}^{hot}$ and $\boldsymbol{\gamma}^{opt}$, across multiple test cases. b)~Scatter plots comparing the bias values  $\boldsymbol{\gamma}^{hot}$ and $\boldsymbol{\gamma}^{opt}$ for various test cases.}
\label{fig:Biases}
\end{figure*}

\begin{figure*}[t!]
\centering
\subfloat[\small ]{
\includegraphics[width=0.485\textwidth]{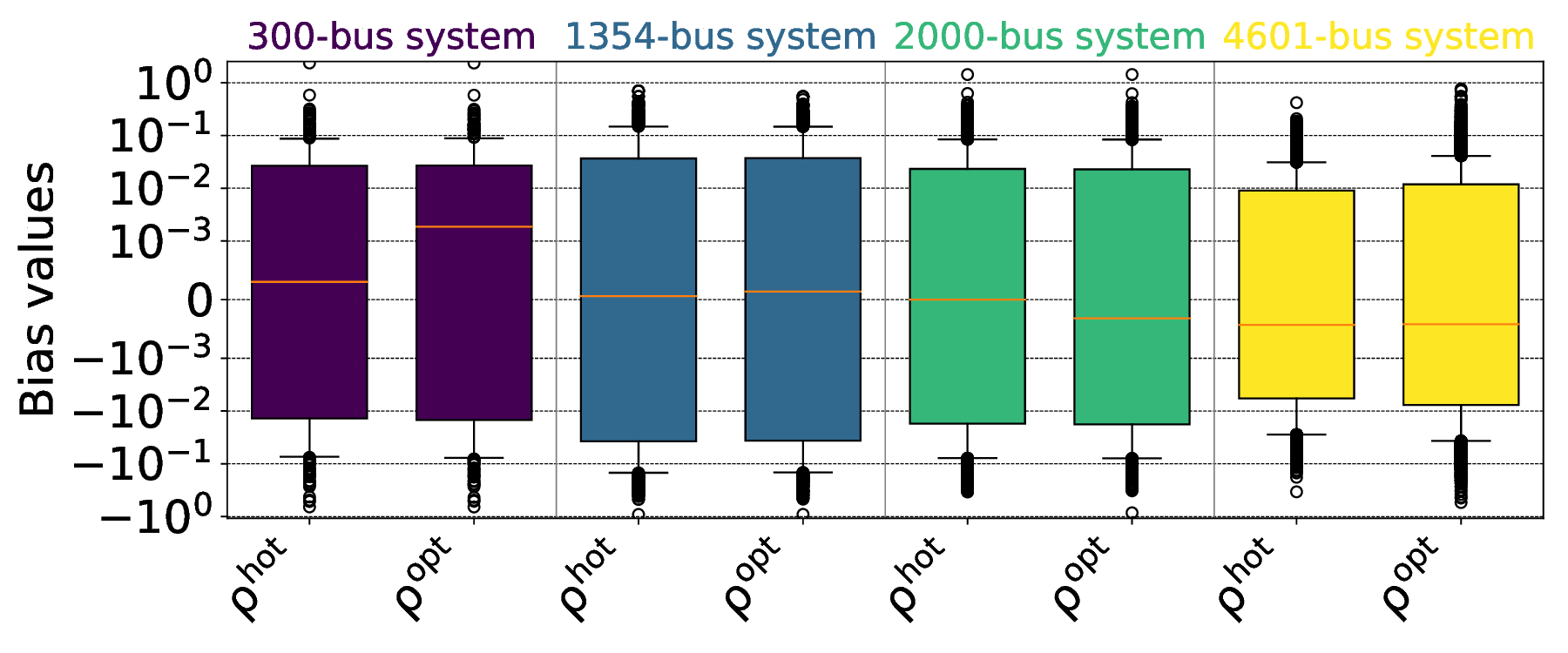}
    \label{fig:boxplot-losses}
}
\hfill
\subfloat[\small  ]{
\includegraphics[width=0.485\textwidth]{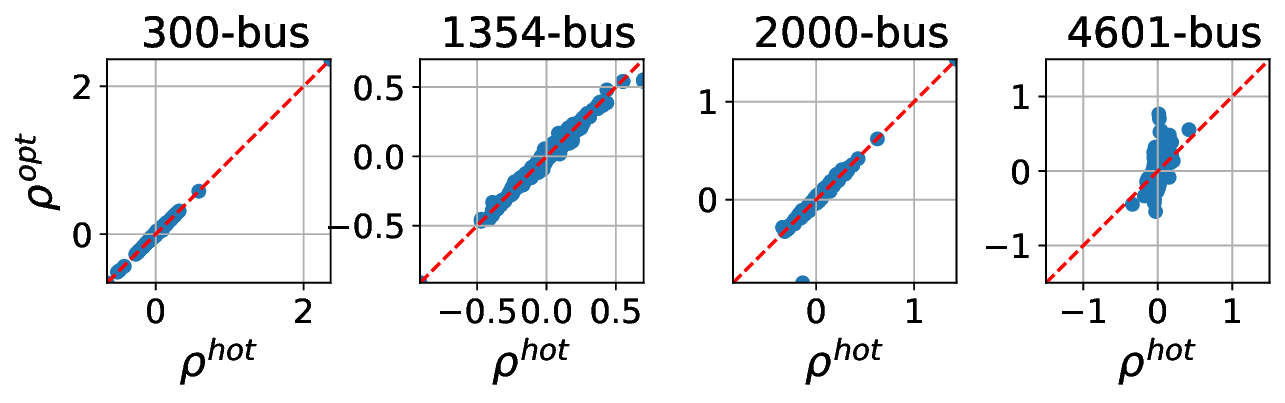}
    \label{fig:losses-scatter}
}
\caption{ (a) Boxplots showing the distributions of hot-start and optimal flow bias values, $\boldsymbol{\rho}^{opt}$ and $\boldsymbol{\rho}^{hot}$, across multiple test cases. (b)~Scatter plots comparing the loss values  $\boldsymbol{\rho}^{hot}$ and $\boldsymbol{\rho}^{opt}$ for various test cases.}
\label{fig:Losses}
\end{figure*}

These boxplots in Fig.~\ref{fig:box-plot} reveal that the distributions of the optimized parameter values align closely with those from existing heuristics for selecting $\mathbf{b}$ (i.e., $\mathbf{b}^{cold}$ and $\mathbf{b}^{cold}_{r=0}$, and $\mathbf{b}^{hot}$). This indicates that our proposed algorithm yields parameter values within a reasonable range, with variations that are consistent with traditional heuristics for choosing $\mathbf{b}$.
Furthermore, Fig.~\ref{fig:coefficient-scatter} provides scatter plots comparing the hot-start coefficient values ($\mathbf{b}^{hot}$) with optimized ones ($\mathbf{b}^{opt}$) across various test cases. The red dashed line at $45^\circ$ in each subplot signifies a one-to-one correlation in the parameter values. Similarly, Figs.~\ref{fig:Biases} and~\ref{fig:Losses} compare the distributions of the optimized bias parameters $\boldsymbol{\gamma}^{opt}$ and  $\boldsymbol{\rho}^{opt}$ with the hot-start DC parameters $\boldsymbol{\gamma}^{hot}$ and  $\boldsymbol{\rho}^{opt}$.
Optimized parameters are broadly similar to those from existing heuristics, suggesting an alignment with longstanding power engineering intuition that the line susceptances are a key parameter in dictating power flows. However, there are some lines for which the optimized $\mathbf{b}$, $\boldsymbol{\gamma}$, and $\boldsymbol{\rho}$ values differ from those in traditional heuristics. These differences suggest that targeted adjustments to the $\mathbf{b}$, $\boldsymbol{\gamma}$, and $\boldsymbol{\rho}$ parameters can substantially improve the DC power flow approximation's accuracy.

%



\subsection{Accuracy with Respect to the AC Power Flow}
\label{subsec:Comparative Analysis}
\begin{figure}[!t]
\centering
\includegraphics[width=3.5in]{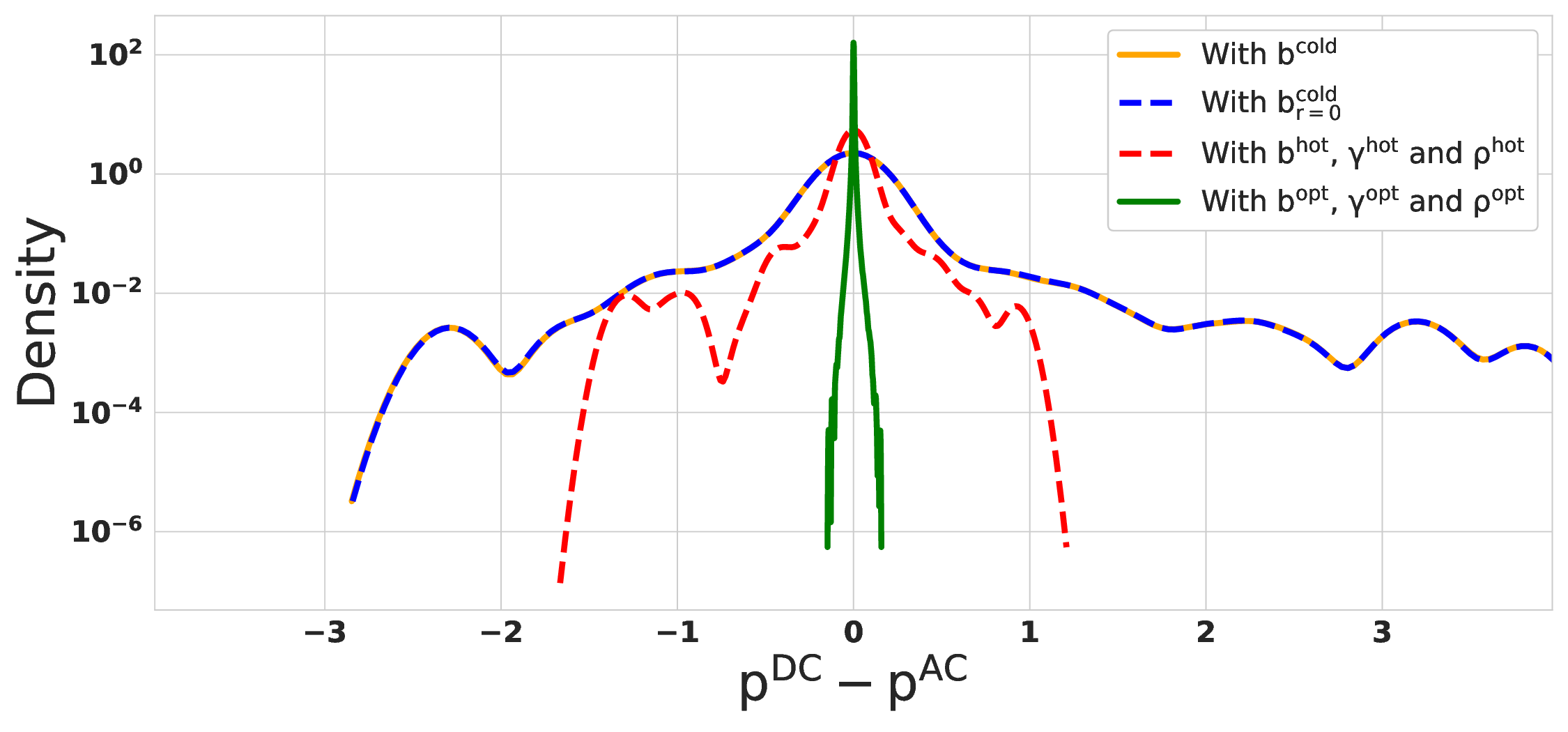}
\caption{Density distributions of the difference between AC and DC power flows in per unit with optimized parameters ($\mathbf{b}^{opt}$, $\boldsymbol{\gamma}^{opt}$, and $\boldsymbol{\rho}^{opt}$), cold-start parameters (both $\mathbf{b}^{cold}$ and $\mathbf{b}^{cold}_{r=0}$, with $\boldsymbol{\gamma}$ and $\boldsymbol{\rho}$ equal to zero), and hot-start parameters ($\mathbf{b}^{hot}$, $\boldsymbol{\gamma}^{hot}$, and $\boldsymbol{\rho}^{hot}$) for the \texttt{Pegase 1354-bus} system over 2000 testing scenarios.}
\label{fig:density_diff}
\end{figure}

\begin{figure}[!t]
\centering
\includegraphics[width=3.5in]{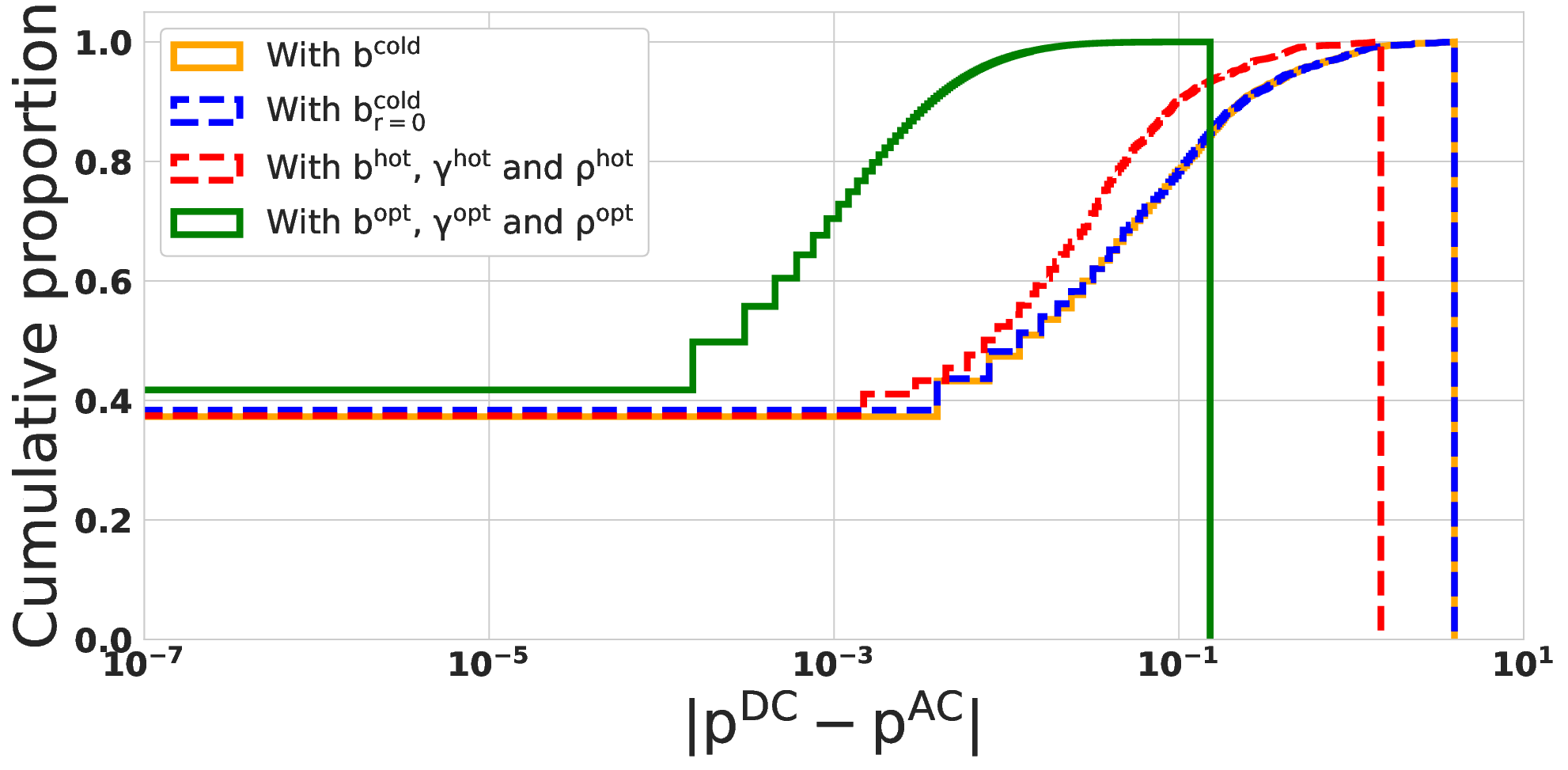}
\caption{Cumulative proportion of the absolute error between AC and DC power flows for the \texttt{Pegase 1354-bus} system. The graph compares four scenarios: usage of cold-start $\mathbf{b}^{cold}$ (orange); $\mathbf{b}^{cold}_{r=0}$ (blue); hot-start $\mathbf{b}^{hot}$, $\boldsymbol{\gamma}^{hot}$, and $\boldsymbol{\rho}^{hot}$ (red); and optimized $\mathbf{b}^{opt}$, $\boldsymbol{\gamma}^{opt}$, $\boldsymbol{\rho}^{opt}$ parameters, shown on a logarithmic scale.}
\label{fig:error_CDF}
\end{figure}

To benchmark the accuracy of our algorithm, we next perform comparisons to the AC power flow model. Fig.~\ref{fig:density_diff} illustrates the density distributions of errors achieved when using the optimized and traditional DC parameters for the \texttt{Pegase 1354-bus} system over 2000 testing scenarios. Fig.~\ref{fig:error_CDF} shows the accuracy advantages of our optimized parameters, with maximum errors less than $0.151$ per unit versus errors up to $3.965$, $3.965$, and $1.484$ per unit resulting from the cold-start DC approximation $\mathbf{b}^{cold}$ and $\mathbf{b}^{cold}_{r=0}$ and the hot-start DC approximation with $\mathbf{b}^{hot}$, $\boldsymbol{\gamma}^{hot}$, and $\boldsymbol{\rho}^{hot}$, respectively.

\begin{table*}\centering
\vspace*{1em}
\smaller
\caption{Squared Two-Norm and $\infty$-Norm Loss Functions for Different Test Cases with the L-BFGS, BFGS, and TNC Methods}
\setlength{\tabcolsep}{1.5pt} 
\renewcommand{\arraystretch}{1.1}
\begin{tabularx}{\textwidth}{p{0.085cm}>{\centering\arraybackslash}p{1.3cm}SSSSSSSSSS>{\arraybackslash}p{1.9cm}}
\toprule
 & \textbf{Test case}& \multicolumn{5}{c}{\textbf{Squared Two-Norm Loss}} & \multicolumn{5}{c}{\textbf{$\infty$-Norm Loss}} & \textbf{Training (s)}\\ 
\cmidrule(lr){3-7} \cmidrule(lr){8-12}
& & $\mathbf{b}^{cold}$ & $\mathbf{b}^{cold}_{r=0}$ & Hot~start& Optimization & Factor~of& $\mathbf{b}^{cold}$ & $\mathbf{b}^{cold}_{r=0}$ &  Hot~start&  Optimization &Factor~of& \\ 
\midrule \midrule
\multirow{6}{*}{\rotatebox{90}{L-BFGS}} 
& \texttt{14-bus} & 2.571 & 2.379 & 0.048 & 0.028 &  {($92, 85, 2)\times$} &0.202 & 0.183 & 0.059 & 0.054 &{($4, 4, 1)\times$} &$\mathbf{3}$ \\
& \texttt{57-bus} &1.471 & 1.743 & 0.033& 0.015 &    {($98, 116, 2)\times$}     & 0.144 & 0.169 &0.057& 0.048 &{($3, 3, 1)\times$}& $\mathbf{5}$ \\
& \texttt{118-bus} & 25.465 & 35.410 &  1.070& 0.078 &{($326, 454, 14)\times$} & 1.182& 1.460  &0.301& 0.169 &{($7, 9, 2)\times$}& 51 \\
& \texttt{200-bus} & 0.215& 0.212  &0.00001&0.00001 &{($21500, 21200, 1)\times$} & 0.130 & 0.130 &0.009& 0.003  &{($43, 43, 3)\times$}& $\mathbf{9}$ \\
& \texttt{300-bus} & 180.437& 177.831  &0.115& 0.026  &{($6940, 6940, 4)\times$} & 4.310 & 4.310 &0.185& 0.194 &{($22, 22, 1)\times$}& $\mathbf{17}$ \\
& \texttt{1354-bus}& 176.081 & 176.400  &28.909& 0.036 &{($4891, 4900, 803)\times$} & 3.964 & 3.964 &1.484& 0.159 &{($25, 25, 9)\times$}& 24896 \\
& \texttt{2000-bus}& 376.248 & 378.497  &59.162& N/A &N/A & 16.328 & 16.328 &6.877& N/A &N/A& 149 \\
& \texttt{4601-bus}& 8.732&8.501    &0.158& N/A &N/A &  2.087& 2.132 & 0.303&N/A  &N/A&  1272\\

\midrule 
\multirow{6}{*}{\rotatebox{90}{BFGS}} 
& \texttt{14-bus} & 2.571 & 2.379 &0.048& \hphantom{000}$\mathbf{0.025}$&{($103, 95, 2)\times$} & 0.202& 0.183  &0.059& \hphantom{000} $\mathbf{0.050}$&{($4, 4, 1)\times$}& 50 \\
& \texttt{57-bus} & 1.471 &1.743 &0.033& 0.015  &{($92, 85, 2)\times$} & 0.144 & 0.169 &0.057& 0.048 &{($3, 3, 1)\times$}&75  \\

& \texttt{118-bus} &  25.465 & 35.410  &1.070& 0.087  &{($293, 407, 12)\times$} & 1.182& 1.460  &0.213& 0.175  &{($7, 8, 1)\times$}& 77 \\
& \texttt{200-bus} & 0.215& 0.212  &  0.00001& 0.00001 &{($21500, 21200, 1)\times$} & 0.130 & 0.130 &0.009 &0.003  &{($43, 43, 3)\times$}& 265 \\

& \texttt{300-bus} & 180.437& 177.831  &0.115& 0.026 &{($6940, 6940, 4)\times$} & 4.310 & 4.310 &0.185& 0.191 &{($22, 22, 1)\times$}& 328 \\

& \texttt{1354-bus}& 176.081& 176.400  &28.909& 0.035  &{($4891, 4900, 803)\times$} & 3.964& 3.964  &1.484& 0.160  &{($25, 25, 9)\times$}& 3412 \\

& \texttt{2000-bus} & 376.248& 378.497  &59.162&\hphantom{00}$\mathbf{0.003}$ & {($125416, 126166, 19721)\times$} &16.328& 16.328 &6.877&  0.171  &{($96, 96, 40)\times$}& 22336 \\
& \texttt{4601-bus}& 8.732&8.501    &0.158& N/A &N/A &  2.087& 2.132 & 0.303&N/A  &N/A&  72\\

\midrule
\multirow{6}{*}{\rotatebox{90}{TNC}} 
& \texttt{14-bus} & 2.571 & 2.379 &0.048& 0.027 &{($95, 88, 2)\times$} & 0.202& 0.183  &0.059& 0.053  &{($4, 4, 1)\times$}& 24 \\
& \texttt{57-bus} & 1.471&1.743  &0.033& \hphantom{00} $\mathbf{0.015}$ &{($98, 116, 2)\times$} & 0.144& 0.169  &0.057& \hphantom{00} $\mathbf{0.047}$  &{($3, 3, 1)\times$}& 22 \\

& \texttt{118-bus} &  25.465& 35.410 & 1.070& \hphantom{00} $\mathbf{0.076}$  &{($335, 466, 14)\times$} & 1.182& 1.460  &0.301& \hphantom{00} $\mathbf{0.158}$ &{($7, 9, 2)\times$}& $\mathbf{38}$ \\
& \texttt{200-bus} & 0.215& 0.212  &0.00001& \hphantom{00000} $\mathbf{0.00001}$ &{($21500, 21200, 1)\times$} & 0.130 & 0.130 &0.009& \hphantom{00}$\mathbf{0.003}$  &{($43, 43, 3)\times$}& $\mathbf{9}$ \\

& \texttt{300-bus} & 180.437& 177.831  &0.115& \hphantom{00}$\mathbf{0.024}$ &{($7518, 7410, 5)\times$} & 4.310 & 4.310 &0.185& \hphantom{00}$\mathbf{0.185}$&{($22, 22, 1)\times$}& 63 \\

& \texttt{1354-bus}  & 176.081& 176.400  &28.909& \hphantom{00}$\mathbf{0.032}$ &{($5502, 5512, 903)\times$} & 3.965& 3.965  &1.484& \hphantom{00}$\mathbf{0.151}$ &{($25, 25, 9)\times$}& $\mathbf{1186}$ \\

& \texttt{2000-bus} & 376.248& 378.497  &59.162& \hphantom{00}$\mathbf{0.003}$ & {($125416, 126166, 19721)\times$} & 16.328 & 16.328 &6.877& \hphantom{00}$\mathbf{0.170}$  &{($96, 96, 40)\times$}& $\mathbf{2993}$ \\

& \texttt{4601-bus} & 8.732& 8.501 &0.158& \hphantom{000}$\mathbf{0.0009}$ &{($9702, 9446, 176)\times$}  & 2.087& 2.132  &0.303& \hphantom{00}$\mathbf{0.058}$& {($36, 37, 5)\times$}&$\mathbf{23154}$ \\

\bottomrule
\end{tabularx}
\begin{tablenotes}
 \item[*] \scriptsize The best performing method (i.e., smallest loss function) is bolded for each test case.
  \item[*] \scriptsize N/A indicates numerical difficulties (the method returns the initial parameters).
  \item[*] \scriptsize The ``Factor of'' columns show the factors of improvements relative to the cold-start model with $\mathbf{b}^{cold}$, the cold-start model with $\mathbf{b}^{cold}_{r=0}$, and the hot-start model with $\mathbf{b}^{hot}$, $\boldsymbol{\gamma}^{hot}$, and $\boldsymbol{\rho}^{hot}$, respectively. 
\end{tablenotes}
\label{table:combined_loss_function_infinity_norm_and_time}
\end{table*}

Table~\ref{table:combined_loss_function_infinity_norm_and_time} provides a detailed comparison of the squared two-norm and $\infty$-norm loss functions evaluated for different test cases using three optimization methods: L-BFGS, BFGS, and TNC. 
While the TNC method exhibits superior performance in many cases, it does not universally outperform the \mbox{L-BFGS} and BFGS methods. For instance, the BFGS method yields better results for certain loss metrics for the \texttt{IEEE 14-bus} case. In addition, the training time for the L-BFGS method is usually much smaller than other methods while having a loss value comparable to the TNC method. 

For every test case, the use of the optimally selected parameters $\mathbf{b}^{opt}$, $\boldsymbol{\gamma}^{opt}$, and $\boldsymbol{\rho}^{opt}$ reduces the loss function values, implying increased accuracy in the DC power flow. The substantial improvements in both the squared two-norm and $\infty$-norm loss demonstrates the effectiveness of our algorithm. For example, applying the TNC method to the \texttt{4601-bus} test case results in the squared two-norm loss decreasing from $0.158$ in hot-start DC model to $0.0009$ (a factor of $176$ improvement) and the $\infty$-norm loss decreasing from $0.303$ to $0.058$ (a factor of $5\times$ improvement). Similar trends are observed across all test cases and optimization methods.

The table also shows the training time (in seconds) for various optimization methods. While the training times increase with the system size, even reasonably large systems (several thousand buses) remain within acceptable times for offline computations (several hours). We expect that further efforts in selecting and tuning optimization methods and more efficient implementations would lead to additional computational improvements for the training process.

We also note that the online execution times required to solve the DC power flow problems with our optimized parameters are comparable to the DC power flow solution times for existing parameter heuristics such as using $b^{cold}$, $b^{cold}_{r=0}$, or hot-start. Specifically, the average per DC power flow solution times with cold-start, hot-start, and optimized values range from $1$ to $931$ milliseconds across the test cases.

\subsection{Application to $N-1$ Contingency Analysis}
\label{sec:Robustness}

Given the unpredictable nature of real-world power systems, the capacity to effectively handle changes in topology is a critical characteristic for any power flow model. In particular, the $N-1$ contingency scenarios, where any single component may fail, are important considerations for power system operations and planning.

There are multiple ways one could handle contingencies. In traditional approaches with $\mathbf{b}^{cold}$ or $\mathbf{b}^{cold}_{r=0}$, one would simply remove any lines outaged in a line contingency scenario from the problem. With our optimization-based algorithm, one could take the same approach by setting the values of $\mathbf{b}^{opt}$, $\boldsymbol{\rho}^{opt}$ corresponding to outaged lines to zero, and adjusting $\boldsymbol{\gamma}^{opt}$ accordingly. Maintaining accurate performance with this approach would suggest that our algorithm generalizes well across related network topologies (i.e., the parameters are not ``overfit'' for a particular topology). Alternatively, we could solve tailored optimization problems for each contingency to find the optimal parameters $\mathbf{b}^{opt}_{tail}$, $\boldsymbol{\gamma}^{opt}_{tail}$, and $\boldsymbol{\rho}^{opt}_{tail}$ specific to each scenario. This requires more training time and memory for computing and storing the many additional parameters that must be selected. However, these computations are trivially parallelizable and thus well suited for a high-performance computing setting since the optimization problems for each contingency scenario can be run without requiring any information from other contingencies. 

To explore these different approaches, we next describe a small-scale experiment using the \texttt{IEEE 14-bus} system. For each line contingency, we optimized the parameters $\mathbf{b}$, $\boldsymbol{\gamma}$, and $\boldsymbol{\rho}$ to minimize the loss function~\eqref{eq:objective_function} while solely considering power injection scenarios corresponding to that contingency. 

The results in Table~\ref{table:contingency} show the performance of the tailored parameters $\mathbf{b}^{opt}_{tail}$, $\boldsymbol{\gamma}^{opt}_{tail}$, and $\boldsymbol{\rho}^{opt}_{tail}$ versus cold-start, hot-start, and base-case-optimized parameters. Tailoring parameters for individual contingency scenarios consistently yields superior results. On average, using the $\mathbf{b}^{opt}_{tail}$, $\boldsymbol{\gamma}^{opt}_{tail}$, and $\boldsymbol{\rho}^{opt}_{tail}$ parameters provides a $98.70\%$ improvement over the cold-start approach. When compared to the hot-start heuristic, $\mathbf{b}^{opt}_{tail}$, $\boldsymbol{\gamma}^{opt}_{tail}$, and $\boldsymbol{\rho}^{opt}_{tail}$ exhibits an improvement of approximately $92\%$. We also note that the $\mathbf{b}^{opt}_{base}$, $\boldsymbol{\gamma}^{opt}_{base}$, and $\boldsymbol{\rho}^{opt}_{base}$ parameters consistently surpass both the cold-start and hot-start approaches across all test scenarios. On average, $\mathbf{b}^{opt}_{base}$, $\boldsymbol{\gamma}^{opt}_{base}$, and $\boldsymbol{\rho}^{opt}_{base}$ parameters show an improvement of approximately $89\%$ over the cold-start approach and approximately $16\%$ over the hot-start approach. This indicates that our proposed algorithm exhibits superior generalizability compared to traditional parameter selection heuristics.

To compute DC power flow approximation parameters that simultaneously consider accuracy with respect to both the base case and contingencies, one could combine base case and contingency scenarios as inputs to the proposed algorithm. However, in our experiments, this combined approach did not yield satisfactory results. We are currently exploring an alternative approach of clustering related contingency scenarios and optimizing parameters specific to each cluster. This allows for more tailored parameter selection while mitigating the computational burden involved in calculating different parameter values for each contingency.

\begin{table}[t]
    \centering
    \caption{Loss function evaluated for different $N-1$ contingencies\\ in the IEEE 14-bus system (20 Branches)}
    \setlength{\tabcolsep}{1.5pt} 
    \renewcommand{\arraystretch}{1.2}
    \begin{tabular}{|c|c|c|c|c|c|}
        \hline
        Contin. & $\mathbf{b}^{cold}$ & $\mathbf{b}^{cold}_{r=0}$ & $(\,\cdot\,)^{hot}$ & $(\,\cdot\,)^{opt}_{base}$  &$(\,\cdot\,)^{opt}_{tail.}$ \\
        \hline
         1 & 39.62 & 39.91 & 24.40 & 23.58  & 1.89 \\
         2 & 7.71 & 7.97 & 2.36 & 2.24  & 0.09 \\
         3 & 7.72 & 6.90 & 2.07 & 1.87 &  0.16 \\
         4 & 3.51 & 3.35 & 0.44 & 0.40 &  0.04 \\
         5 & 3.02 & 2.75 & 0.50 & 0.48 &  0.04 \\
         6 & 2.69 & 2.22 & 0.40 & 0.30 &  0.04 \\
         7 & 3.55 & 3.40 & 0.22 & 0.18 &  0.07 \\
         8 & 2.71 & 2.41 & 0.07 & 0.03 &  0.03 \\
         9 & 2.66 & 2.39 & 0.09 & 0.04 &  0.03 \\
        10 & 3.47 & 3.31 & 0.24 & 0.24 &  0.03 \\
        11 & 2.63 & 2.39 & 0.28 & 0.27 &  0.04 \\
        12 & 2.66 & 2.46 & 0.17 & 0.14 &  0.03 \\
        13 & 3.16 & 2.72 & 0.77 & 0.71 &  0.03 \\
        14 & 2.70 & 2.50 & 0.09 & 0.04 &  0.03 \\
        15 & 2.93 & 2.57 & 0.19 & 0.09 &  0.04 \\
        16 & 2.49 & 2.33 & 0.19 & 0.17 &  0.03 \\
        17 & 2.64 & 2.50 & 0.42 & 0.38 &  0.03 \\
        18 & 2.55 & 2.35 & 0.08 & 0.08 &  0.03 \\
        19 & 2.58 & 2.38 & 0.08 & 0.06 &  0.03 \\
        20 & 2.63 & 2.41 & 0.17 & 0.17 &  0.03 \\

        \hline
    \end{tabular}

    \begin{tablenotes}
 \item[*] \qquad \qquad $(\,\cdot\,)$ stands for $\mathbf{b}$, $\boldsymbol{\gamma}$, and $\boldsymbol{\rho}$ parameters.
\end{tablenotes}
    \label{table:contingency}
\end{table}


\section{Conclusion}
\label{sec:Conclusion}

This paper presents a machine learning-inspired algorithm to improve the DC power flow approximation's accuracy by optimizing the selection of the coefficient and bias parameters. Our algorithm harnesses L-BFGS, BFGS, and TNC optimization methods to refine the coefficient and bias parameters, achieving better agreement between the DC and AC power flow models.
Our simulations on various test systems demonstrate the effectiveness of this algorithm. We improve the accuracy of the DC power flow approximation by several orders of magnitude across a range of test cases. These findings underline the value of our algorithm in enhancing the reliability and accuracy of the DC power flow model, particularly for large-scale power systems.

Our future work intends to focus on applying the improved DC power flow model to several critical applications in power systems, such as optimal power flow, unit commitment, and optimal transmission switching. We anticipate that the accuracy gained from our enhanced DC power flow model could lead to significantly improved performance in these and other applications. 
Regarding next steps in contingency analyses, our experiments showed that naively combining base case and contingency scenarios was not effective. Moving forward, we are focusing on a clustering approach: grouping related contingency scenarios and optimizing parameters for each cluster. This strategy aims to optimize performance across diverse scenarios without overfitting.
Our ongoing work also aims to reduce training time. This may involve targeted scenario sampling and methods inspired by techniques for accelerating the training of machine learning models.

\bibliographystyle{IEEEtran}
\bibliography{refs}

\appendix[Derivation of Sensitivities]
\label{sec:sensitivity_derivation}

The sensitivities of the $\mathbf{p}^{DC}$ with respect to $\mathbf{b}$ are calculated using \eqref{eq:coef_sensitivity}, which is derived in this appendix.

Starting from the initial function \eqref{eq:branchPowerFlow2}, we have:
\begin{equation}
\label{eq:start}
\mathbf{p}^{DC} = \text{diag}(\mathbf{b})  \mathbf{A} [\mathbf{A}^T \text{diag}(\mathbf{b})  \mathbf{A}]^{-1}(\mathbf{P}-\boldsymbol{\gamma}) +\boldsymbol{\rho}.
\end{equation}
For notational convenience, let $\mathbf{M} = \mathbf{A} [\mathbf{A}^T \text{diag}(\mathbf{b})  \mathbf{A}]^{-1}(\mathbf{P}-\boldsymbol{\gamma})$. Now, \eqref{eq:start} can be rewritten as $\mathbf{p}^{DC} =\text{diag}(\mathbf{b}) \mathbf{M}$. We then differentiate with respect to $\mathbf{b}$:
\begin{equation}
d\mathbf{p}^{DC} = d(\text{diag}(\mathbf{b})) \mathbf{M} + \text{diag}(\mathbf{b}) d\mathbf{M}.
\end{equation}
To compute the differential of $\mathbf{M}$ with respect to $\mathbf{b}$, we differentiate the expression for $\mathbf{M}$, which yields:
\begin{equation}
\begin{aligned}
d\mathbf{M} =& - \mathbf{A} [\mathbf{A}^T \text{diag}(\mathbf{b})  \mathbf{A}]^{-1} d(\mathbf{A}^T \text{diag}(\mathbf{b})  \mathbf{A}) \\
&\qquad \qquad \qquad \qquad \times [\mathbf{A}^T \text{diag}(\mathbf{b})  \mathbf{A}]^{-1}(\mathbf{P}-\boldsymbol{\gamma}).
\end{aligned}
\end{equation}

\vfill\eject
Here, the differential of the product $\mathbf{A}^T \text{diag}(\mathbf{b})  \mathbf{A}$ with respect to $\mathbf{b}$ is $d(\mathbf{A}^T \text{diag}(\mathbf{b})  \mathbf{A}) = \mathbf{A}^T d(\text{diag}(\mathbf{b})  \mathbf{A})$. Substituting this into the previous equation yields:
\begin{equation}
\begin{aligned}
d\mathbf{M} =& - \mathbf{A} [\mathbf{A}^T \text{diag}(\mathbf{b})  \mathbf{A}]^{-1} \mathbf{A}^T d(\text{diag}(\mathbf{b})  \mathbf{A}) \\
& \qquad \qquad \qquad \qquad  \times[\mathbf{A}^T \text{diag}(\mathbf{b})  \mathbf{A}]^{-1}(\mathbf{P}-\boldsymbol{\gamma}).
\end{aligned}
\end{equation}

Substituting the expression for $d\mathbf{M}$ back into the expression for $d\mathbf{p}^{DC}$, we get:
\begin{equation}
\begin{aligned}
d\mathbf{p}^{DC} &= d(\text{diag}(\mathbf{b})) \mathbf{M} - (\text{diag}(\mathbf{b})  \mathbf{A}) [\mathbf{A}^T \text{diag}(\mathbf{b})  \mathbf{A}]^{-1} \\&  \times
\mathbf{A}^T d(\text{diag}(\mathbf{b}) ) \mathbf{A} [\mathbf{A}^T \text{diag}(\mathbf{b})  \mathbf{A}]^{-1}(\mathbf{P}-\boldsymbol{\gamma}).
\end{aligned}
\end{equation}

This equation can be simplified by vectorizing both sides. We apply the vectorization operator, denoted as $\text{vec}(\,\cdot\,)$ to obtain:
\begin{equation}
\begin{aligned}
\textrm{vec}(d\mathbf{p}^{DC}) &=   
 \textrm{vec} (d(\text{diag}(\mathbf{b})) \mathbf{M}) \\
  &- \textrm{vec} \Big( \text{diag}(\mathbf{b})  \mathbf{A} [\mathbf{A}^T \text{diag}(\mathbf{b})  \mathbf{A}]^{-1}  \mathbf{A}^T \\ & \times 
  d(\text{diag}(\mathbf{b}) ) \mathbf{A}[\mathbf{A}^T \text{diag}(\mathbf{b})  \mathbf{A}]^{-1}(\mathbf{P}-\boldsymbol{\gamma}) \Big).
\end{aligned}
\end{equation}

We can vectorize this equation as:
\begin{equation}
\begin{aligned}
\text{vec}(d\mathbf{p}^{DC}) &=  \Big(\mathbf{M}^T \otimes \mathbf{I} \Big) \text{vec}(d(\text{diag}(\mathbf{b}))) - \Big(\mathbf{M}^T \otimes  (\text{diag}(\mathbf{b})\\ &
 \times \mathbf{A} [\mathbf{A}^T \text{diag}(\mathbf{b})  \mathbf{A}]^{-1} \mathbf{A}^T \Big)\text{vec}(d(\text{diag}(\mathbf{b}))),
\end{aligned}
\end{equation}
where $\otimes$ stands for the Kronecker product. Isolating the derivative of $\text{vec}(\mathbf{p}^{DC})$ with respect to $\text{vec}(\text{diag}(\mathbf{b}))$ yields:
\begin{equation}
\begin{aligned}
\mathbf{\Psi} =& \frac{ \text{vec}(d \mathbf{p}^{DC})}{ \text{vec}(d (\text{diag}(\mathbf{b})))} = \mathbf{M}^T \otimes \Big(\mathbf{I} - (\text{diag}(\mathbf{b})  \mathbf{A} \\& \qquad \qquad \qquad \qquad \times[\mathbf{A}^T \text{diag}(\mathbf{b})  \mathbf{A}]^{-1} \mathbf{A}^T) \Big),
\end{aligned}
\end{equation}
where $\mathbf{\Psi}$ is the matrix representing the sensitivity of each line flow $\mathbf{p}^{DC}$ to changes in each parameter $\text{diag}(\mathbf{b})$. Finally, we only need the sensitivity of the line flows $\mathbf{p}^{DC}$ with respect to the diagonal elements of $\text{diag}(\mathbf{b})$, i.e., $\mathbf{b}$, which can be calculated as follows:
\begin{align}
\label{eq:sensitivity2}
\frac{\partial \mathbf{p}^{DC}}{\partial \mathbf{b}} = & \text{~diag} (\mathbf{M}^T) \times \Big (\mathbf{I} - \text{diag}(\mathbf{b}) \mathbf{A} [\mathbf{A}^T \text{diag}(\mathbf{b}) \mathbf{A}]^{-1} \mathbf{A}^T \Big).
\end{align}

\end{document}